\documentclass[twocolumn]{jpsj2}

\def\runauthor{Takahiro Yamamoto, Rumi Manago, Yoshihiro Mori and Chikara Ishii}

\title{%
Cusp Singularities in the Magnetization Curve of Frustrated Kondo Necklace Model
}

\author{%
Takahiro~Yamamoto \thanks{E-mail address: takahiro@rs.kagu.tus.ac.jp},
Rumi~Manago, Yoshihiro~Mori and Chikara~Ishii
}

\inst{%
Department of Physics, Faculty of Science, Tokyo University of Science, 
1-3 Kagurazaka, Shinjuku-ku, Tokyo 162-8601, Japan
}

\recdate{\today}

\abst{%
Magnetization processes of frustrated Kondo
necklace model are studied by means of a density matrix renormalization 
group (DMRG) method and an elementary band theory based on a bond-operator 
formalism. The DMRG calculations clearly show the cusp singularity
in a low-magnetization region ($0<m<1/2$) besides that in a high-magnetization 
region ($1/2<m<1$) which is expected from previous studies on the magnetization 
curve of the Majumdar-Ghosh model.
An appearance mechanism of the low-magnetization cusp is interpreted in terms of 
a double-well shape of a low-energy band arising from 
frustrations between nearest- and next-nearest-neighbor interactions.
We also discuss critical behaviors 
of magnetization near the cusp and obtain a phase diagram showing 
whether the cusp appears in the magnetization curve or not.   
}

\kword{%
magnetization curve, cusp singularity, frustration, DMRG, bond-operator
}

\begin{document}
\maketitle

\section{\label{sec:level1}Introduction}
Over the last few years, a large number of materials exhibiting quantum phase 
transitions induced by an external magnetic field have been discovered, and 
their critical properties have been attracted much attention experimentally 
as well as theoretically. Such field-induced phase transitions are associated 
with various singularities in the magnetization curve, 
such as plateaus~\cite{ref:Hida,ref:Totsuka,ref:oshi,ref:shira},
cusps~\cite{ref:Parkinson,ref:Kiwata,ref:Schmitt-M-K,ref:Kiwata-A,ref:Schlott,ref:Zvyagin,ref:fujii} 
and jumps~\cite{ref:Kohno-T,ref:ger,ref:hira}. In contrast to concern 
about plateaus and jumps, very few attempts have been made at the problem 
of cusp singularity in the magnetization curve. The cusp singularity was 
first found in the SU(3) integrable spin chain known as the Lai-Sutherland 
model by Parkinson with the use of the Bethe ansatz method~\cite{ref:Parkinson}.
Subsequently, it has been investigated with a focus on integrable models.
This may be because it is difficult to find cusp singularities of non-integrable 
models by conventional numerical approaches like an exact-diagonalization and quantum
Monte Calro methods.

Very recently, Okunish {\it et al}. showed that the cusp singularities appear in 
the magnetization curve of non-integrable spin systems, such as the spin-$1/2$ 
zig-zag ladder and the spin-$1$ bilinear-biquadratic chain, using the 
product-wave-function renormalization group (PWFRG) method,~\cite{ref:oku1,ref:oku2} 
which is a variant of the density-matrix renormalization group (DMRG) 
method~\cite{ref:white}. They also pointed out that the appearance of cusps 
originate in the double-well shape of the energy dispersion of the quasiparticles 
such as magnon, spinon excitations.

In this paper, we introduce {\it frustrated Kondo necklace model}, 
which is equivalent to the spin-$1/2$ chain with nearest- and 
next-nearest-neighbor interactions ($J_1$ and $J_2$) coupled to 
a localized spin on each site, as a new non-integrable spin system 
exhibiting the cusp singularities in its magnetization curve. 
Here, let us briefly review the magnetization process of the 
frustrated Kondo necklace model already understood by previous 
studies~\cite{ref:yama1,ref:yama2,ref:yama3}. The low-energy 
excitations of this system have a finite gap (so-called Kondo singlet gap),
reflecting that the ground state in the absence of the magnetic field 
is always in a quantum disordered phase whose spin configuration is 
an aggregate of intra-atomic Kondo singlets~\cite{ref:zhang,ref:sca,ref:mou,ref:otsu,ref:chen}.  
When a magnetic field, $h$, is applied to the system, the magnetization of 
this system begins to grow at the Kondo singlet gap. As for the Kondo 
necklace model without $J_2$~\cite{ref:doni}, the magnetization curve 
smoothly grows with increasing the magnetic field, $h$, once the Kondo 
singlet gap is collapsed~\cite{ref:yama1}. On the other hand, in the 
presence of the strong frustration ($J_2/J_1>0.3419$), the magnetization 
plateau appears at half saturation ($m=1/2$) besides the Kondo singlet 
gap~\cite{ref:yama2,ref:yama3}. The $m=1/2$ plateau can be explained as 
follows. The frustrated Kondo necklace model with $m=1/2$ can be effectively 
regarded as spin-$1/2$ zig-zag ladder which is well known as the Majumdar-Ghosh 
model~\cite{ref:Majumdar-G} in a special case $J_2/J_1=0.5$, because the 
innercore spins of the system are almost ferromagnetically aligned by 
the magnetic field $h$ and their degree of freedom is almost frozen by 
$h$. It is well known that the ground state of spin-$1/2$ zig-zag ladder 
is in the quantum disordered singlet-dimer phase. This phase corresponds 
to the $m=1/2$ plateau phase in the present system.

Since the magnetization curve of the spin-$1/2$ zig-zag ladder exhibits 
the cusp singularity in the vicinity of the saturation magnetization,
we can expect that the magnetization curve of the present model exhibits 
a similar cusp singularity in the high-field region $1/2<m<1$. The expected 
cusp will be found in the high-field region with the use of DMRG method in 
the next section. Moreover, the DMRG calculations will find an unexpected cusp
just above the Kondo singlet gap. An appearance mechanism of the low-field 
cusp differs from that of the high-field cusp, because the excitations of 
innercore spins play an essential role of the magnetization process in this 
region. The main purpose of this paper is to explain the origin 
and characteristics of the low-field cusp singularity.

This paper is organized as follows. In Sec.~\ref{sec:level2}, the Hamiltonian 
of the frustrated Kondo necklace model will be given and its magnetization 
curves will be calculated by the DMRG method. In Sec.~\ref{sec:level3}, the 
frustrated Kondo necklace model will be mapped onto a fermionic quasiparticle 
system by means of the bond-operator representation which can explicitly describe 
the triplet excitation from the Kondo singlet sea (Sec.~\ref{sec:level3A}).
A mechanism of appearance of the low-field cusp will be explained in 
terms of the double-well shape of an energy dispersion curve of the triplet 
excitation (Sec.~\ref{sec:level3B}). Moreover we will discuss the critical 
behavior of the magnetization curve near the cusp singularity in the low-field 
region $0<m<1/2$ (Sec.~\ref{sec:level3C}). In Sec.~\ref{sec:level4}, we will 
determine the phase diagram on a plane of the Kondo interaction, $J_K$, and the 
frustration parameter, $\lambda\equiv J_2/J_1$, showing whether the cusp 
singularity appears in the magnetization curve or not. The final section 
(Sec.~\ref{sec:level5}) is devoted to conclusion.

\section{\label{sec:level2}Hamiltonian and magnetization curves}
\begin{figure}[t]
  \begin{center}
    \includegraphics[keepaspectratio=true,width=70mm]{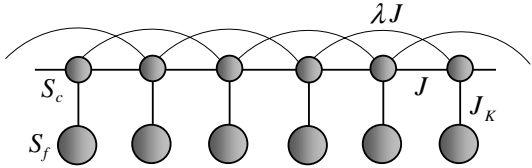}
  \end{center}
    \caption{Schematic representation of the frustrated Kondo 
    necklace model~(\ref{eq:ham}). Small and Large circles denote 
    the conduction-electron spin ($S_c$) and the innercore spin 
    ($S_f$) respectively.}
    \label{fig:1}
\end{figure}

The Kondo necklace model is simplified version of the Kondo lattice model in 
one dimension, assuming a narrow band limit for conduction electrons in the Kondo 
lattice model. In this limit, the motion of conduction electrons is 
prohibited, and the system is composed of two different kinds of localized spins: 
the spins of innercore electrons (on the $f$ orbits) and the conduction electrons. 
Thus, the Hamiltonian of frustrated Kondo necklace is described as
\begin{eqnarray}
{\mathcal H}=J\!\!\sum_{\scriptstyle j=1 \atop \scriptstyle \nu=1,2}^{L}
\lambda^{\nu-1}{\mbox{\boldmath $S$}}_{c,j}\cdot{\mbox{\boldmath $S$}}_{c,j+\nu}
+J_{\rm K}\sum_{j=1}^{L}{\mbox{\boldmath $S$}}_{c,j}\cdot{\mbox{\boldmath $S$}}_{f,j},
\label{eq:ham}
\end{eqnarray}
which is schematically shown in Fig.~\ref{fig:1}.
Here, ${\mbox{\boldmath $S$}}_{c,j}$ are conduction-electron spins with $S_{c,j}=1/2$,
and ${\mbox{\boldmath $S$}}_{f,j}$ are innercore-electron spins with $S_{f,j}=1/2$ which 
are hanging from the $S_{c}$-spin chain. The nearest-neighbor exchange coupling $J$ 
and Kondo coupling $J_K$ are assumed to be antiferromagnetic ($J, J_K>0$). $\lambda$ is 
a frustration parameter which denotes competition between nearest- and next-nearest-neighbor 
interactions in $S_{c}$-spins.

In the calculation of the magnetization, we use the DMRG method based on the 
finite-size algorithm~\cite{ref:white}. The advantage of this method is that 
the large system can be studied, as compared with the exact-diagonalization 
method. Unlike the quantum Monte Carlo simulations, the DMRG method is free 
from the negative sign problem and can easily calculate even the ground state 
in the magnetic field. The DMRG method is an ideal tool for studying the 
magnetization curve of the one-dimensional spin systems. Our DMRG calculations 
have been performed under the open boundary condition, for the system length 
$L=60$ and the number of optimal states up to $120$.

\begin{figure}[t]
  \begin{center}
    \includegraphics[keepaspectratio=true,width=80mm]{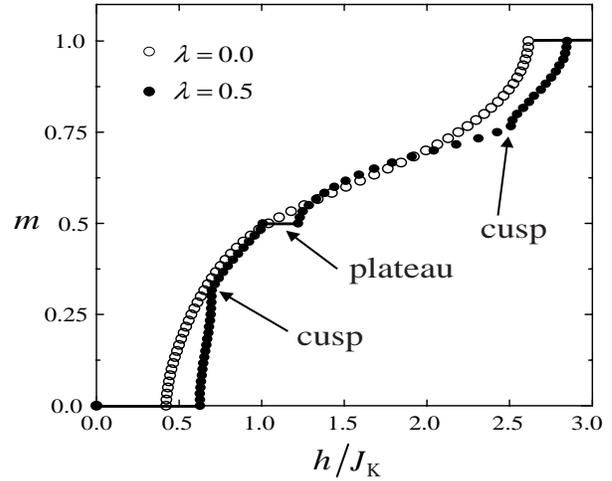}
  \end{center}
  \caption{The magnetization curves of 
  the frustrated Kondo necklace model with $J/J_K=1.0$.
  In the case of $\lambda=0.5$ (filled circle), 
  two cusps and a plateau appear in 
  the magnetization curve which is smooth in 
  the absence of frustration $\lambda=0.0$
  (open circle).}
\label{fig:2}
\end{figure}

Figure~\ref{fig:2} shows the magnetization curves of the frustrated Kondo 
necklace model with $\lambda=0.0$ and $0.5$ for fixed $J/J_K=1.0$. 
In the absence of frustration ($\lambda=0.0$), the magnetization curve grows smoothly 
with increasing the magnetic field $h$ once the Kondo singlet gap $\Delta_K$ collapses. 
This means that there is no phase transition in the magnetic field in the range 
$\Delta_K<h<h_S$ ($h_S$: saturation field) and the system belongs to a universality 
class of the Tomonaga-Luttinger liquid phase~\cite{ref:yama2,ref:yama3}.

In the case of $\lambda=0.5$, the Fig.~{\ref{fig:2}} clearly shows the cusp singularities 
in the low-field ($0<m<1/2$) and high-field ($1/2<m<1$) regions, which cannot be detected 
by the exact-diagonalization calculations. As we mentioned in Sec.~\ref{sec:level1}, 
the appearance of high-field cusp is expected from previous results by Okunishi 
{\it et al.}~\cite{ref:oku1}, and can be interpreted in terms of the low-energy 
dispersion with double-well shape of single down-spin magnon in the up-spin sea.
However, the concept of single down-spin magnon is useful only in the vicinity of 
the saturation magnetization. Therefore it cannot be used to explain an appearance 
mechanism of the low-field cusp. To discuss the low-field cusp, we must consider 
appropriate elementary excitations in the low-field region.

\section{\label{sec:level3} Mechanism of an appearance of a cusp singularity}
\subsection{\label{sec:level3A}Bond-operator representation}
In this section, we discuss an origin of cusp singularity in the low-field region 
($0<m<1/2$). First of all, it should be remarked that we have to know about 
the low-energy excitation modes of the system, in order to understand the magnetization 
process in the low-field region. As was stressed before, the ground state of the present 
system (\ref{eq:ham}) in the absence of the magnetic field is always in the Kondo singlet 
phase, when the Kondo coupling, $J_K$, of any strength is present. This suggests that the 
low-energy excitations will be well described by triplet excitations from the Kondo singlet 
sea. To express this picture in an explicit form, we introduce the following bond-operators 
creating singlet and triplet states out of the vacuum $\left|{0}\right\rangle$:
\begin{eqnarray}
s^\dagger\left|0\right\rangle
=\frac{1}{\sqrt{2}}\left(\left|\uparrow\Downarrow\right\rangle
-\left|\downarrow\Uparrow\right\rangle\right),
\label{eq:s}\\
t_x^\dagger\left|0\right\rangle
=-\frac{1}{\sqrt{2}}\left(\left|\uparrow\Uparrow\right\rangle
-\left|\downarrow\Downarrow\right\rangle\right),
\label{eq:t_x}\\
t_y^\dagger\left|0\right\rangle
=\frac{i}{\sqrt{2}}\left(\left|\uparrow\Uparrow\right\rangle
+\left|\downarrow\Downarrow\right\rangle\right),
\label{eq:t_y}\\
t_z^\dagger\left|0\right\rangle
=\frac{1}{\sqrt{2}}\left(\left|\uparrow\Downarrow\right\rangle
+\left|\downarrow\Uparrow\right\rangle\right).
\label{eq:t_z}
\end{eqnarray}
These bond-operators were first proposed by Sachdev and Bhatt~\cite{ref:sach} in order to 
treat the quantum phase transitions between dimerized and magnetically ordered phases of 
antiferromagnets. Here, $\uparrow,\downarrow$ and $\Uparrow,\Downarrow$ denote $z$-components 
of $S_c$ and $S_f$ spins, respectively. To describe the magnetization process of the Kondo 
necklace model, it is more convenient to use another set of bond-operators $u^\dagger$ and 
$d^\dagger$ creating the states $\left|\uparrow\Uparrow\right\rangle$ and 
$\left|\downarrow\Downarrow\right\rangle$:
\begin{eqnarray}
u^\dagger\left|0\right\rangle
\equiv \frac{-1}{\sqrt{2}}\left(t_x^\dagger+it_y^\dagger\right)\left|0\right\rangle
=\left|\uparrow\Uparrow\right\rangle,
\label{eq:up}\\
d^\dagger\left|0\right\rangle
\equiv \frac{1}{\sqrt{2}}\left(t_x^\dagger-it_y^\dagger\right)\left|0\right\rangle
=\left|\downarrow\Downarrow\right\rangle,
\label{eq:down}
\end{eqnarray} 
instead of the transverse triplet operators $t_x$ and $t_y$~\cite{ref:wang}.

In terms of these bond-operator representation, we can imagine that each lattice point of 
the Kondo necklace model is in one of the four atomic states
$s^\dagger\left|{0}\right\rangle$, $u^\dagger\left|{0}\right\rangle$,
$t_z^\dagger\left|{0}\right\rangle$, and $d^\dagger\left|{0}\right\rangle$,
or equivalently, each site is occupied by one of the virtual particles
created by bond-operators $s^\dagger$, $u^\dagger$, $t_z^\dagger$ and $d^\dagger$.
Then, these operators should satisfy the local constraint,
\begin{eqnarray}
s_j^\dagger s_j+u_j^\dagger u_j+t_{z,j}^{\dagger}t_{z,j}+d_j^{\dagger}d_j=1,
\label{eq:constraint}
\end{eqnarray}
in order to work in the physical subspace with a single particle on each site.
The spin operators $S_{p,j}$ ($p=c,f$) are represented in terms of these bond-operators as
\begin{eqnarray}
S_{p,j}^{+}&=&\frac{1}{\sqrt{2}}\left(
\pm s_j^{\dagger}d_{j}\mp u_{j}^{\dagger}s_j+t_{z,j}^{\dagger}d_{j}+u_{j}^{\dagger}t_{z,j}
\right),
\label{eq:S+}\\
S_{p,j}^{-}&=&\frac{1}{\sqrt{2}}\left(
\pm d_{j}^{\dagger}s_j\mp s_j^{\dagger}u_{j}+d_{j}^{\dagger}t_{z,j}+t_{z,j}^{\dagger}u_{j}
\right),
\label{eq:S-}\\
S_{p,j}^{z}&=&\frac{1}{2}\left(
\pm s_j^{\dagger}t_{z,j}\pm t_{z,j}^{\dagger}s_j-d_{j}^{\dagger}d_{j}+u_{j}^{\dagger}u_{j}
\right),
\label{eq:Sz}
\end{eqnarray}
where the signs in the upper and lower rows denote the cases of $p=c$ and $f$, respectively.

Thus far, we have not specified the algebra of the bond-operators. As a matter of fact, 
the spin algebra:
\begin{eqnarray}
\left[ S_{p,i}^{+},S_{q,j}^{-}\right]=2S_j^{p,z}\delta_{i,j}\delta_{p,q},\quad
\left[ S_{p,i}^{\pm},S_{q,j}^{z}\right]=\mp S_j^{p,\pm}\delta_{i,j}\delta_{p,q},
\end{eqnarray}
can be derived by using the expressions~(\ref{eq:S+})-(\ref{eq:Sz}), irrespective that 
the bond-operators obey either bosonic or fermionic commutation relations. It is usual 
to assume for the bond-operators to be bosonic, in order to perform a physically meaningful 
mean-field approximation on the basis of Bose-Einstein condensation of singlet states,
$\left\langle s\right\rangle=\left\langle s^\dagger\right\rangle\neq 0$,
$\left\langle t\right\rangle=\left\langle t^\dagger\right\rangle =0$.
The bond-operator mean-field theory was successfully applied to the study of disordered 
dimerized phases in various quantum spin systems (e.g. spin-$1/2$ ladder~\cite{ref:gopa},
Bilayer Heisenberg model~\cite{ref:matsu} and standard Kondo necklace model~\cite{ref:zhang}).
Recently, the method of bosonic bond-operators was applied to the studies of
magnetization process of spin-$1$ antiferromagnetic Heisenberg chain and spin-$1/2$ and 
spin-$1$ Heisenberg ladders, mapping onto spin-$1/2$ pseudospin Hamiltonian in the effective 
magnetic field~\cite{ref:wang}.

In closing this section, we note that the algebra of the bond-operators is not important 
in the following discussion. An important point in this work is that the triplet excitations 
from the Kondo singlet sea are represented as {\it fermionic} particles, as shown in next 
section. Thus, we impose here the bosonic commutation relation on the bond-operators, only 
for convenience. Of course, even if we impose the fermionic commutation relation to the 
bond-operators, our results in this paper are not changed.

\subsection{\label{sec:level3B}Effective band theory}
We are now in the position to state a reasonable picture of the origin of the cusp
singularity in the low-field region, $0<m<1/2$, on the basis of an elementary band 
theory constructed with the use of bond-operator representation. To do this, the 
triplet excitations from the Kondo singlet sea are treated as {\it fermionic} particles.

Since the ground state is always in the Kondo singlet phase as mentioned before,
it seems to be useful to start from the strong Kondo-coupling limit, $J/J_K\ll 1$.  
In the limit of strong-$J_K$ and $h\approx \Delta_K$, we can project out the states 
$t_z^{\dagger}\left|{0}\right\rangle$ and $d^{\dagger}\left|{0}\right\rangle$, 
preserving only two low-lying states $s^{\dagger}\left|{0}\right\rangle$ and 
$u^{\dagger}\left|{0}\right\rangle$. Then, the constraint (\ref{eq:constraint}) 
can be rewritten as
\begin{eqnarray}
s_j^\dagger s_j+u_j^\dagger u_j=1,
\label{eq:constraint2}
\end{eqnarray}
and the spin operators in Eqs.~(\ref{eq:S+})-(\ref{eq:Sz}) are simplified as
\begin{eqnarray}
S_{p,j}^{+}&=&
\mp\frac{1}{\sqrt{2}}u_{j}^{\dagger}s_j\equiv
\mp\frac{1}{\sqrt{2}}\gamma_j^\dagger,
\label{eq:S++}\\
S_{p,j}^{-}&=&
\mp\frac{1}{\sqrt{2}}s_j^{\dagger}u_{j}\equiv
\mp\frac{1}{\sqrt{2}}\gamma_j,
\label{eq:S--}\\
S_{p,j}^{z}&=&
\frac{1}{2}u_{j}^{\dagger}u_{j}\equiv\frac{1}{2}\gamma_j^{\dagger}\gamma_j.
\label{eq:Szz}
\end{eqnarray}
The new operator $\gamma_j^\dagger\equiv u_j^{\dagger}s_j$ creates a triplet particle on 
$j$-th site from the Kondo singlet sea, satisfying the following {\it hardcore bosonic} 
commutation relations
\begin{eqnarray}
\big\{\gamma_j, \gamma_j^\dagger\big\}&\equiv&
\gamma_j\gamma_j^\dagger+\gamma_j^\dagger\gamma_j=1,
\label{eq:hardcore}\\
(\gamma_j^\dagger)^2&=&\gamma_j^2=0,\\
\big[\gamma_i, \gamma_j^\dagger \big]&=&
\big[\gamma_i^\dagger, \gamma_j^\dagger \big]=
\big[\gamma_i, \gamma_j \big]=0,\quad i\neq j.
\end{eqnarray}
Note that the operatots $\gamma_j^\dagger$ and $\gamma_j$ are neither fermionic nor bosonic, 
and thanks to Eq.~(\ref{eq:hardcore}), the constraint (\ref{eq:constraint2}) is satisfied 
automatically. The hardcore bosons can be mapped onto spinless fermions through the
Jordan-Wigner transformation,
\begin{eqnarray}
\xi_j^\dagger&=&\gamma_j^{\dagger}
\exp\left[{\rm i}\pi\sum_{n=1}^{j-1}\gamma_n^{\dagger}\gamma_n\right],\\
\xi_j&=&
\exp\left[-{\rm i}\pi\sum_{n=1}^{j-1}\gamma_n^{\dagger}\gamma_n\right]\gamma_j.
\end{eqnarray}
It is convenient to introduce the Fourier transform of fermionic operator,
\begin{eqnarray}
\xi_j=\frac{1}{\sqrt{L}}\sum_{k}\xi_{k}\exp\left({\rm i}kj\right),
\end{eqnarray}
where the wave number $k$ takes discrete values $k=2\pi{n}/L$ ($n=0,1,\cdots,L-1$) assuming 
$L$ to be even integer. Using the fermionic operators $\xi_{k}^{\dagger}$ and $\xi_k$, the 
frustrated Kondo necklace model [Eq.~(\ref{eq:ham})] in the magnetic field $h$ is mapped onto 
a {\it fermionic} quasiparticle system:
\begin{eqnarray}
{\mathcal H}&=&
E_{0}+\sum_{k}\left(\epsilon_{k}-h\right)\xi_{k}^{\dagger}\xi_{k}\nonumber\\
& &+\frac{J}{4}\sum_{k,k',q}\cos\left(k+k'+q\right)
\xi_{k+q}^{\dagger}\xi_{k'-q}^{\dagger}\xi_{k}\xi_{k'},
\label{eq:ham_k}
\end{eqnarray} 
where $E_0=-3J_{K}L/4$ is an energy of the Kondo singlet sea,
\begin{eqnarray}
\left|{\rm vac}\right\rangle
=s_1^\dagger s_2^\dagger\cdots s_L^\dagger\left|0\right\rangle,
\label{eq:0}
\end{eqnarray}
satisfying a relation $\xi_k\left|{\rm vac}\right\rangle=0$ and the single-particle energy 
dispersion is given as
\begin{eqnarray}
\epsilon_k=J_K+\frac{J}{2}\left[\cos{k}+\lambda\cos\left(2k\right)\right].
\label{eq:disp}
\end{eqnarray}
The last term in Eq.~(\ref{eq:ham_k}) describes the scattering between 
the quasiparticles. This is negligible in the low-density region 
$\sum_{k}\left\langle\xi_k^{\dagger}\xi_k\right\rangle\ll{1}$,
corresponding to the low-magnetization $m\ll{1}$.

\begin{figure}[t]
  \begin{center}
    \includegraphics[keepaspectratio=true,width=70mm]{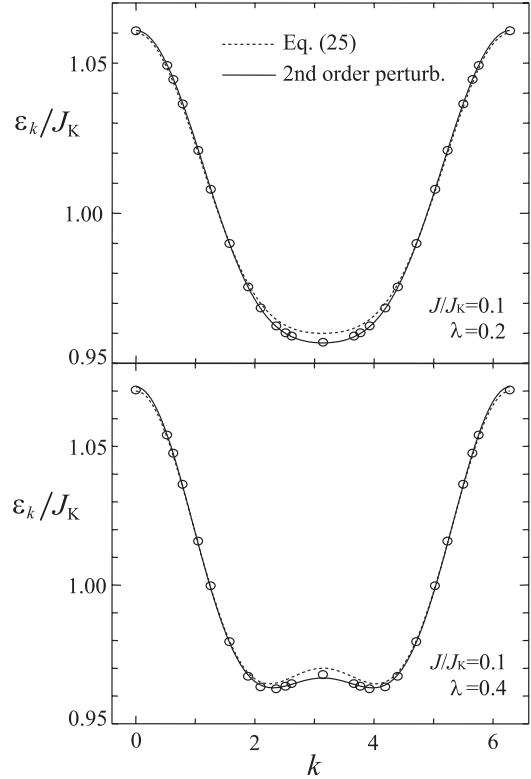}
  \end{center}
  \caption{The energy dispersion of a single triplet fermion for
  $J/J_K=0.1$ with $\lambda=0.2$ (above) and $0.4$ (below).
  Short-dash curve shows the dispersion given by Eq.~(\ref{eq:disp}).
  Solid curve represent the dispersion obtained by 
  the perturbation expansion up to second order in $J/J_K$ (see Appendix).
  The open circles show the exact-diagonalization data at 
  $k=2\pi n/L$ ($n=0,1,\cdots,L-1$) for $L=8$, $10$ and $12$.}
\label{fig:3}
\end{figure}

\begin{figure*}[t]
  \begin{center}
    \includegraphics[keepaspectratio=true,width=140mm]{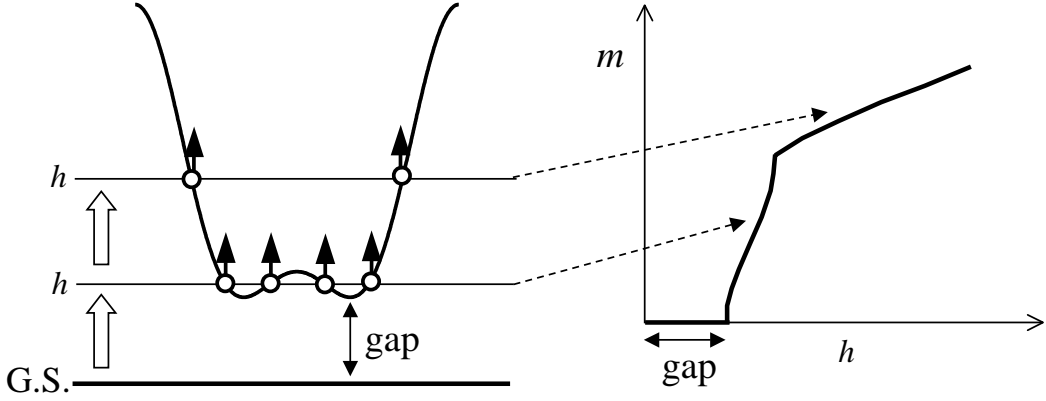}
  \end{center}
  \caption{Schematic diagram showing the relation between the double-well shaped
  low-energy dispersion of triplet quasiparticle and the cusp singularity in the 
  low-field magnetization curve. `G.S.' denotes the energy of ground state.}
\label{fig:4}
\end{figure*}

We discuss the single-particle energy dispersion (\ref{eq:disp}) in some detail.
The dashed curves in Fig.~\ref{fig:3} show the single-particle energy
dispersion $\epsilon_k/J_K$ for $\lambda=0.2$ and $0.4$ where $J/J_K=0.1$ 
is assumed.
As is seen in Fig.~\ref{fig:3} and Eq.~(\ref{eq:disp}), the shape of dispersion
curve changes at the critical frustration $\lambda_c=1/4$.
(Taking account of higher order corrections, this value of the critical point
$\lambda_c=1/4$ will be modified later.) 
For $\lambda<1/4$, the dispersion has a single bottom at $k=\pi$. 
On the other hand, for $\lambda>1/4$, 
the dispersion curve has double-well shape with two bottoms
at the wave numver $\pi\pm\sin^{-1}(-1/4\lambda)$ ($0\le k<2\pi$)
and at the energy $\epsilon_b/J_K=1-(J/2J_K)(\lambda+1/8\lambda)$,
The state at $k=\pi$, being global minimum for $\lambda<1/4$,
turns out to be local maximum with the energy $\epsilon_m/J_K=1-(J/2J_K)(\lambda-1)$.

These results can be also obtained by the perturbation expansion
up to the first order in $J/J_K$, 
and provide a sufficient basis to interpret the origin of the
appearance of the cusp qualitatively.
More detailed analysis up to the second order in $J/J_K$ is described in
the Appendix, and the results are shown by solid curves in Fig.~\ref{fig:3}. 
As is seen from these plots, the second order results are in excellent agreement
with those of exact diagonalization indicated by open circles. 

The eigenstate with the energy $\epsilon_k$ is 
given as a single quasiparticle state,
\begin{eqnarray}
\left|k\right\rangle&=&\xi_{k}^{\dagger}\left|{\rm vac}\right\rangle\nonumber\\
&=&\frac{1}{\sqrt{L}}\sum_{j=1}^{L}\exp\left({\rm i}kj\right)
s_1^\dagger s_2^\dagger\cdots u_{j}^\dagger\cdots s_L^\dagger\left|0\right\rangle.
\label{eq:k}
\end{eqnarray}
To study the low-energy properties, one often makes the assumption that 
additional many quasiparticle eigenstates can be constructed as
a collection of independent quasiparticles, that is 
$\left|k_1,k_2,\cdots,k_N\right\rangle=
\xi_{k_1}^{\dagger}\xi_{k_2}^{\dagger}\cdots\xi_{k_N}^\dagger\left|0\right\rangle$
with the energy $\epsilon_{k_1}+\epsilon_{k_2}+\cdots\epsilon_{k_N}$.

We have prepared now to interpret the low-field magnetization process 
just above the Kondo singlet gap $\Delta_K$ in terms of elementary band theory
with the use of above single-particle energy dispersion relation (\ref{eq:disp}).
Thereby, the magnetic field $h$ plays a role of the chemical potential or 
the Fermi energy and the magnetization $m$ corresponds to the quasiparticle 
number per site occupying the state $\left|k\right\rangle$ below the Fermi energy $h$.
Thus the magnetization at zero temperature is given as
\begin{eqnarray}
m=\frac{1}{L}\sum_{|k|<k_F}\left\langle\xi_{k}^{\dagger}\xi_{k}\right\rangle
\to\frac{1}{2\pi}\int_{0}^{2\pi}\theta(h-\epsilon_k)dk,
\label{eq:m}
\end{eqnarray}
where $k_F$ and $\theta(x)$ are the Fermi wave-number and the Heaviside's step function, 
respectively.

On the basis of Eq.~(\ref{eq:m}), we can explain the characteristic features of 
magnetization curve in the low-field region, shown in Fig.~\ref{fig:4}. When a 
magnetic field $h$ is applied to the system, no magnetization appears until
$h$ arrives at the bottoms of the dispersion. If the frustration parameter 
$\lambda$ is larger than the critical value $\lambda_c$ in the region 
$\epsilon_m>h>\epsilon_b$, the Fermi level crosses the dispersion at four 
wave-numbers, satisfying $\epsilon_k=h$, and the magnetization curve shows
sharp rise. With further increase in the magnetic field, the Fermi level eventually 
arrives at the local maximum $\epsilon_m$ at $k=\pi$, and the number of crossing 
points of the Fermi level and the dispersion curve suddenly reduces from 4 to 2, 
resulting in a sudden slowdown of the rate of band filling and growth rate of the 
magnetization. This sudden change in the rate of band filling is reflected  as the 
cusp singularity on the magnetization curve of the Kondo necklace model.

\subsection{\label{sec:level3C}Critical behavior near the cusp singularity}
\begin{figure}[t]
  \begin{center}
    \includegraphics[keepaspectratio=true,width=80mm]{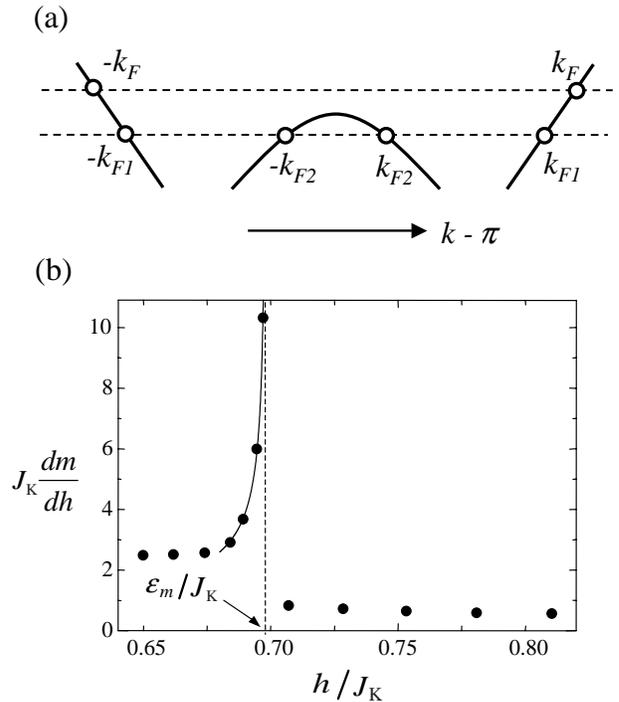}
  \end{center}
  \caption{(a) The effective low-energy dispersion.
  The dashed lines show the magnetic field $h$ corresponding to the Fermi level. 
  (b) The field derivative of the magnetization $dm/dh$ near 
  the critical point $h=\epsilon_m$ of 
  the frustrated Kondo necklace model with $J/J_K=1.0$ and $\lambda=0.5$.
  The solid line denotes asymptotic curve 
  $dm/dh\sim (\epsilon_m-h)^{-1/2}$ near $h=\epsilon_m$.}
  \label{fig:5}
\end{figure}

To discuss the detailed behavior of the magnetization curve in the vicinity of 
cusp singularity, we consider the field derivative of the magnetization $dm/dh$.
From Eq.~(\ref{eq:m}), we find that $dm/dh$ is given by the quasiparticle density 
of states at the Fermi energy $h$ as follows. 
\begin{eqnarray}
\frac{dm}{dh}&=&\frac{1}{2\pi}\int_{0}^{2\pi}\delta(h-\epsilon_k)dk\nonumber\\
&=&\frac{1}{2\pi}\sum_{\{\epsilon_k=h\}}\left|\frac{d\epsilon_k}{dk}\right|^{-1}.
\label{eq:dos} 
\end{eqnarray}
Here, $\sum_{\{\epsilon_k=h\}}$ denotes a sum over all Fermi points. When the energy 
dispersion is shaped like the double-well, two different situations may occur 
(see Fig.~\ref{fig:5}(a)).

\vspace{5mm}
[1] For $h>\epsilon_m$, there are two Fermi points $\pm k_F$, and
the energy dispersion can be linerized around the Fermi points as far as 
low-energy excitations are concerned: 
$\epsilon_k=\pm v_F\{(k-\pi)\mp k_F\}+h$, where 
$v_F$ is the Fermi velocity of the $\xi_k$-fermions.
Thus, from Eq.~(\ref{eq:dos}), we obtain $dm/dh=L/\pi v_F$ which 
is almost independent of $h$ because $v_F$ only depends weakly on $h$.   

\vspace{5mm}
[2] For $h<\epsilon_m$, there are four Fermi points satisfying the relation,
$2k_{F1}-2k_{F2}=m$. In this case, low-energy states are effectively described by
the two-band model [See Fig.~\ref{fig:5}(a)]:
\begin{eqnarray}
\epsilon_k^{(1)}&=&\pm v_{F1}\{(k-\pi)\mp k_{F1}\}-h,
\label{eq:omega1}\\
\epsilon_k^{(2)}&=&-C(k-\pi)^2+\epsilon_m,
\label{eq:omega2} 
\end{eqnarray}
where $C$ is a positive constant.
The parabolic dependence of $\epsilon_k^{(2)}$ around $k=\pi$ gives 
decisive difference in $dm/dh$ from the case of [1].
Substituting Eqs.~(\ref{eq:omega1}) and (\ref{eq:omega2}) for Eq.~(\ref{eq:dos}), 
we obtain the most dominant contribution
\begin{eqnarray}
\frac{dm}{dh}\sim \frac{L}{2\pi\sqrt{C}}\frac{1}{\sqrt{\epsilon_m-h}}
\end{eqnarray}
which diverges at $h=\epsilon_m$.
Thus, we find that the critical exponent characterizing
the intensity of the divergence of the field derivative of the magnetization, 
$dm/dh$, is given as $1/2$. \\

Plots in Fig.~\ref{fig:5}(b) show $dm/dh$ of the frustrated Kondo necklace 
model (\ref{eq:ham}) with $J/J_K=1.0$ and $\lambda=0.5$, obtained by numerically 
differentiating the DMRG data in Fig.~\ref{fig:2}. As is expected from the above 
results, for $h>\epsilon_m$, the plots depend on $h$ very weakly. On the other hand, 
for $h<\epsilon_m$, it diverges as $dm/dh\sim(\epsilon_m-h)^{-1/2}$ as $h\to \epsilon_m$ 
(the solid curve in Fig.~\ref{fig:5}(b)). Away from the cusp singular point $h=\epsilon_m$, 
the data are almost independent of $h$. This means that the low-energy dispersion is 
described by two linear-bands involving four Fermi points. 
From the above results, we find that the magnetization behaves as 
\begin{eqnarray}
m-m_{\rm cusp}\propto\left\{
  \begin{array}{cc}
     h-\epsilon_m          & :\quad h>\epsilon_m \\
     -\sqrt{\epsilon_m-h}  & :\quad h<\epsilon_m \\
  \end{array}
\right.
\end{eqnarray}
in the vicinity of the critical point $h=\epsilon_m$ with $m=m_{\rm cusp}$.
This result well explains the behavior of magnetization curves near 
the cusp singularity obtained by the DMRG calculations (See Fig.\ref{fig:2}).

\section{\label{sec:level4}Phase Diagram}
In the previous sections, we saw that the appearance of low-field cusp singularity 
in the magnetization curve is governed by the competition between the Kondo coupling 
$J/J_K$ and frustration parameter $\lambda$. In the strong Kondo coupling limit 
$J/J_K\to 0$, the cusp singularity appears when $\lambda$ exceeds the value 
$\lambda_c=1/4$. In the presence of Kondo coupling of general strength, 
the value of $\lambda_c$ will depend on the strength of Kondo coupling. In the 
following, we determine a phase boundary on the plane of $J/J_K$ and 
$\lambda$, dividing whether the cusp appears in the magnetization curve or not. 
To do this, we note that the value of energy dispersion curve at zone-center point 
$k=\pi$ turns from global minimum to local maximum. Therefore, the critical point 
$\lambda_c$ is to be determined from the condition that the point $k=\pi$ is the 
deflection point of the energy dispersion:
\begin{eqnarray}
\left(\frac{\partial^2\epsilon_k}{\partial k^2}\right)_{k=\pi}=0.
\label{eq:cond}
\end{eqnarray}
As shown in Appendix, the critical value of frustration $\lambda_c$, 
including the second order correction with respect to $J/J_K$ is given as
\begin{eqnarray}
\lambda_c=\frac{2J_K+2J}{8J_K+5J}\approx\frac{1}{4}
+\frac{3}{32}\left(\frac{J}{J_K}\right)
-\frac{15}{256}\left(\frac{J}{J_K}\right)^2.
\label{eq:lambda_c}
\end{eqnarray}
In order to determine $\lambda_c$ outside the strong Kondo-coupling region, 
however, we need the expression of energy dispersion 
including the full effects of Kondo coupling and frustration. 
This is performed by determining the energy dispersion for various values of
$J/J_K$ by means of the exact-diagonalization method.

Here we briefly explain our exact-diagonalization method. 
Our exact-diagonalization calculations were performed under 
the periodic boundary condition $S_{p,j}=S_{p,j+L}$ ($p=c$ and $f$).
Since the $z$-component of total spin operator commutes with 
the frustrated Kondo necklace Hamiltonian (\ref{eq:ham}), 
all eigenvalues and eigenstates of the Hamiltonian are labeled by 
the total magnetization $M=\sum_{j=1}^L\left(\sigma_j^z+S_j^z\right)$.
In addition, since the Hamiltonian (\ref{eq:ham}) is 
invariant under the translation along the chain-direction 
($S_{p,j}\to S_{p,j+1}$), 
the eigenvalues and eigenstates are classified by 
the wave number $k=2\pi n/L$ ($n=0,1,\cdots L-1$) 
corresponding to eigenvalue of the translation operator.
Thus, in Fig.~\ref{fig:3}, we plot the energy dispersion, 
$E(k,M=1)-E(k,M=0)$ with $k=2\pi n/L$, for three lengths 
$L=8$, $10$, and $12$.

\begin{figure}[t]
  \begin{center}
    \includegraphics[keepaspectratio=true,width=80mm]{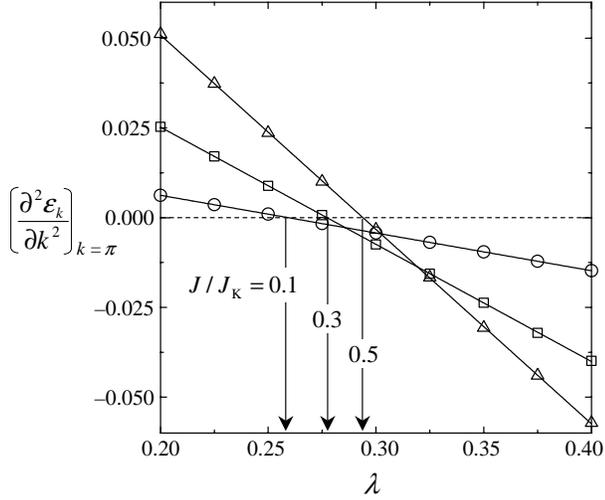}
  \end{center}
  \caption{The second derivative of the energy dispersion $\epsilon_k$ at $k=\pi$.
  The circles, squares and triangles denote 
  $\left(\partial^2\epsilon_k/\partial k^2\right)_{k=\pi}$
  for $J/J_K=0.1$, $0.3$ and $0.5$, respectively. 
  Each symbol is fitted by a straight line.}
\label{fig:6}
\end{figure}

\begin{figure}[t]
  \begin{center}
    \includegraphics[keepaspectratio=true,width=75mm]{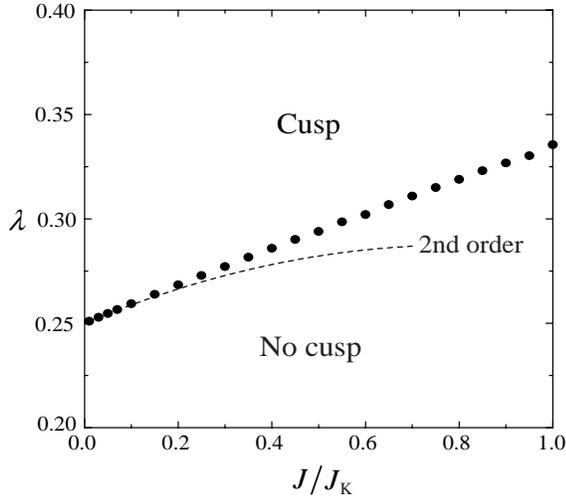}
  \end{center}
  \caption{The phase diagram on plane of $J/J_K$ and $\lambda$, which 
  decides whether a cusp appears in the region of $0<m<1/2$
  in the magnetization curve.
  The short-dash curve is the boundary given by Eq.~(\ref{eq:lambda_c}) 
  obtained from the strong Kondo-coupling expansion up to second order. 
  }
\label{fig:7}
\end{figure}

Since the energy dispersion is an even function around $k=\pi$,
we fit the exact-diagonalization data near $k=\pi$ using a fitting function,  
\begin{eqnarray}
\epsilon_k=\Delta_K+\frac{\alpha}{2!}(k-\pi)^2+\frac{\beta}{4!}(k-\pi)^4+\cdots,
\label{eq:fit}
\end{eqnarray}
up to eighth order and estimate fitting parameters $\Delta_K$, $\alpha$, $\beta, \cdots$
for various $\lambda$ and fixed $J/J_K$.
The parameter  $\alpha=(\partial^2\omega/\partial k^2)_{k=\pi}$ 
thus obtained for $J/J_K=0.1$, $0.3$ and $0.5$
are shown by circles, squares and triangles in Fig.~\ref{fig:6}, 
respectively.
The next step is to interpolate $\alpha$ in order to 
obtain the critical point $\lambda_c$ satisfying the condition (\ref{eq:cond}).
We assumed a liner form of interpolation function $\alpha=A(\lambda-\lambda_c)$
in analogy with that in the strong Kondo-coupling limit (see Appendix).
Thus we can obtain the critical point $\lambda_c$ from 
the condition $\alpha=0$ (see Fig.~\ref{fig:6}).  
Finally, the phase diagram on ($J/J_K,\lambda$)-plane is constructed,
performing above series of procedures for various values of $J/J_K$, 
as shown in  Fig.~\ref{fig:7}. 
The short-dash curve in Fig.~\ref{fig:7} shows the boundary given by 
Eq.~(\ref{eq:lambda_c}) and is in good agreement with the exact-diagonalization 
data in the strong Kondo-coupling region, $J/J_K\ll 1$.
On the boundary, the magnetization curve grows rapidly at $h=\Delta_K$ because 
the low-energy part of the dispersion becomes quartic. 
To be more explicit, the magnetization behaves as $m\sim (h-\Delta_K)^{3/4}$, 
as is derived from Eq.~(\ref{eq:dos}).

\section{\label{sec:level5}Conclusion}
In this paper, the magnetization processes of frustrated Kondo necklace model
have been studied by using the density matrix renormalization group (DMRG) method.
We have paid attention to the low-field cusp singularity appearing just above the 
Kondo singlet gap. To reveal an appearance mechanism of the cusp, we introduced the 
triplet quasiparticle excited from Kondo singlet sea by using the bond-operator 
representation of dimerized spins, and we have discussed the mechanism in terms of 
an elementary band theory with the use of a single-particle energy dispersion of the 
triplet quasiparticle. As a result, we have made it clear that the appearance of 
low-field cusp has its origin in a change of shape of quasiparticle energy dispersion 
curve, reflecting effects of frustration.

We also discussed the critical behavior of the magnetization in the vicinity of 
the cusp singularity in the low-field region.
Calculating the density of states from the quasiparticle energy dispersion, 
we found that the magnetization curve behaves as 
$m-m_{\rm cusp}\propto -\sqrt{\epsilon_{m}-h}$ ($h<\epsilon_{m}$)
and
$m-m_{\rm cusp}\propto h-\epsilon_m$ ($h>\epsilon_{m}$). 
This result well explains the behavior of the magnetization curves near the cusp
obtained by the DMRG calculations. 
Moreover, we obtained the phase diagram on the $(J/J_K, \lambda)$-plane 
deciding whether the cusp appears in the magnetization curve or not.

The effective band theory based on the bond-operator formalism
must be successfully applied to various quantum spin systems with 
dimer-gapped ground state~\cite{ref:yama4}.

\begin{acknowledgments}
We are grateful to M.~Asano for his assistance in the DMRG calculations. We also would 
like to thank Profs. M.~Suzuki, K.~Hida, T.~Nikuni, M.~Nakamura and K.~Okunishi for 
valuable comments and useful discussions. This work has been supported in part by Frontier Research Center for 
Computational Science (FRCCS) of Tokyo University of Science.
\end{acknowledgments}

\appendix

\section{Strong Kondo coupling expansion up to second order in $J/J_K$}
In this appendix, we present the second order correction in $J/J_K$ to 
the low-energy dispersion of the frustrated Kondo necklace model.

The Kondo coupling Hamiltonian [the second term in Eq.~(\ref{eq:ham})] will be 
treated as the unperturbed one and its ground state is given as a direct product 
of the Kondo singlets [Eq.~(\ref{eq:0})]. On the other hand, intra-chain Hamiltonian 
[the first term in Eq.~(\ref{eq:ham})] will be treated as the perturbation. The first 
order perturbation energy vanishes owing to the orthogonal condition. The second-order 
correction to the energy of ground state is calculated by the standard perturbation 
formula. The perturbation energy up to the second order in $J/J_K$ is obtained as
\begin{eqnarray}
E_0/J_K=-\frac{3L}{4}\left\{1+\frac{1}{8}(1+\lambda^2)\left(\frac{J}{J_K}\right)^2\right\}.
\label{eq:A1}
\end{eqnarray}

To obtain the low-energy dispersion, we next calculate the energy correction in 
an excited state. We suppose that the low-energy physics in the strong Kondo-coupling 
region is well described by the single triplet excitation state (\ref{eq:k}). Following 
the same procedure as calculating the ground-state energy, we can estimate the excitation 
energy. The first and the second order energies are given as 
\begin{eqnarray}
E_{k}/J_K&=&\left(-\frac{3}{4}L+1\right)+
\frac{J}{2J_K}\left(\cos k+\lambda\cos(2k)\right)\nonumber\\
& &-\frac{1}{32}\left(\frac{J}{J_K}\right)^2\big\{(1+\lambda^2)(3L+2)\nonumber\\
& &-4(2-\lambda)\cos(k)+2(1-4\lambda^2)\cos(2k)\nonumber\\
& &+4\lambda\cos(3k)+2\lambda^2\cos(4k)\big\}.
\label{eq:A2}
\end{eqnarray}
Subtracting Eq.~(\ref{eq:A2}) from Eq.~(\ref{eq:A1}), we obtain the single-particle 
energy dispersion up to the second order in $J/J_K$:
\begin{eqnarray}
\epsilon_k/J_K&\equiv&(E_{k}-E_0)/J_K\nonumber\\
&=&1-\frac{1}{16}\left(\frac{J}{J_K}\right)^2(1+\lambda^2)\nonumber\\
& &+\frac{J}{8J_K}\left\{4+(2-\lambda)\frac{J}{J_K}\right\}\cos k\nonumber\\
& &+\frac{J}{16J_K}\left\{8\lambda-(1-4\lambda^2)\right\}\frac{J}{J_K}\cos(2k)\nonumber\\
& &-\frac{\lambda}{8}\left(\frac{J}{J_K}\right)^2
\cos(3k)-\left(\frac{\lambda J}{4J_K}\right)^2\cos(4k),
\label{eq:A3}
\end{eqnarray}
where the first order correction is consistent with Eq.~(\ref{eq:disp}). The 
coefficients $\Delta_K$, $\alpha$ and $\beta$ of the Taylor expansion [Eq.~(\ref{eq:fit})] of the 
energy dispersion (\ref{eq:A3}) are given by
\begin{eqnarray}
\Delta_K&=&1-\frac{J}{8J_K}\left(1-\lambda\right)
\left[4+(3+\lambda)\frac{J}{J_K}\right],\\
\alpha&=&-\frac{J}{J_K}\left(2+\frac{5J}{4J_K}\right)
\left(\lambda-\frac{2J_K+2J}{8J_K+5J}\right),
\label{eq:A4}\\
\beta&=&\frac{J}{J_K}\left[8\lambda-\frac{1}{2}-
\frac{J}{4J_K}\left(48\lambda^2-41\lambda+5\right)\right].
\end{eqnarray}
Using the expression for $\alpha$, we can determine an asymptotic curve of 
boundary on the plane of $J/J_K$ and $\lambda$, dividing whether the cusp 
appears in the magnetization curve or not. Since the boundary, $\lambda_c$, 
is determined by the condition, $\alpha=0$, we can obtain the boundary form 
(\ref{eq:lambda_c}) as shown in Fig.~\ref{fig:7}.

\end{document}